\documentclass[aps,prl, twocolumn,floatfix,superscriptaddress,preprintnumbers,tightenlines,showpacs,showkeys,nofootinbib,notoccite]{revtex4-1}
\usepackage[utf8]{inputenc}
\usepackage[colorlinks=true,citecolor=blue,linkcolor=blue]{hyperref}
\usepackage[normalem]{ulem}
\usepackage{amsmath,amssymb, mathrsfs}
\usepackage{epsfig}
\usepackage{graphicx}               
\usepackage{url}
\usepackage{color}
\usepackage{slashed}
\usepackage{multirow}
\usepackage{placeins}
\usepackage[dvipsnames]{xcolor}
\usepackage{epstopdf}
\usepackage{soul}
\usepackage{tikz}
\usepackage[capitalise, english]{cleveref}
\usepackage{siunitx}
\usepackage{xspace}
\usepackage{booktabs}
\usetikzlibrary{trees}
\usetikzlibrary{decorations.pathmorphing}
\usetikzlibrary{decorations.markings}
\usepackage{physics}

\newcommand\myshade{80}
\colorlet{mylinkcolor}{ForestGreen}
\colorlet{mycitecolor}{Red}
\colorlet{myurlcolor}{violet}

\hypersetup{
  linkcolor  = mylinkcolor!\myshade!black,
  citecolor  = mycitecolor!\myshade!black,
  urlcolor   = myurlcolor!\myshade!black,
  colorlinks = true
}

\definecolor{jblue}{RGB}{20,50,100}
\definecolor{npurple}{RGB} {153, 51, 204}
\definecolor{wred}{RGB}{217,0,56}
\definecolor{white}{RGB}{255,255,255}

\definecolor{korange}{RGB}{235, 80,  43}
\definecolor{korange2}{RGB}{245, 100,  63}
\definecolor{kyelloworange}{RGB}{255, 210,  110}
\definecolor{kyelloworange2}{RGB}{240, 170,  90}
\definecolor{kred}{RGB}{204,  102, 153}
\definecolor{kpurple}{RGB}{153,  61, 190}
\definecolor{kpurplelight}{RGB}{213,  161, 230}

 \definecolor{tobycolour}{rgb}{.5,.0,.5}

\DeclareSIUnit\year{yr}
\DeclareSIUnit\pc{pc}
\DeclareSIUnit\ergs{ergs}
\DeclareSIUnit\msun{\ensuremath{M_\odot}}
\allowdisplaybreaks

\makeatletter
\providecommand*{\diff}%
  {\@ifnextchar^{\DIfF}{\DIfF^{}}}
\def\DIfF^#1{%
  \mathop{\mathrm{\mathstrut d}}%
    \nolimits^{#1}\gobblespace}
\def\gobblespace{%
  \futurelet\diffarg\opspace}
\def\opspace{%
  \let\DiffSpace\!%
  \ifx\diffarg(%
    \let\DiffSpace\relax
  \else
    \ifx\diffarg[%
      \let\DiffSpace\relax
    \else
        \ifx\diffarg\{%
        \let\DiffSpace\relax
      \fi\fi\fi\DiffSpace}

\usepackage{tikz,xcolor,hyperref}
\newcommand{\cevns}{CE$\nu$NS\xspace}
\newcommand{\eV}{\text{e\kern-0.15ex V}\xspace}
\newcommand{\keV}{\text{k\eV}\xspace}
\newcommand{\MeV}{\text{M\eV}\xspace}
\newcommand{\GeV}{\text{G\eV}\xspace}

\definecolor{lime}{HTML}{A6CE39}
\DeclareRobustCommand{\orcidicon}{\hspace{-1mm}
	\begin{tikzpicture}
	\draw[lime, fill=lime] (0,0) 
	circle [radius=0.16] 
	node[white] {{\fontfamily{qag}\selectfont \tiny \,ID}};
	\draw[white, fill=white] (-0.0525,0.095) 
	circle [radius=0.007];
	\end{tikzpicture}
	\hspace{-3mm}
}

\foreach \x in {A, ..., Z}{\expandafter\xdef\csname orcid\x\endcsname{\noexpand\href{https://orcid.org/\csname orcidauthor\x\endcsname}
			{\noexpand\orcidicon}}
}

\newcommand{\keep}[1]{\textcolor{blue}{ #1}}

\usepackage{ulem}
\usepackage{xcolor}      

\keywords{}

\begin{document}

\title{First constraint on the weak mixing angle using direct detection experiments}

\author{Tarak Nath Maity\orcidA{}}
\email{tarak.maity.physics@gmail.com}
\affiliation{School of Physics, The University of Sydney and ARC Centre of Excellence for Dark Matter Particle Physics, NSW 2006, Australia}

\author{C\'eline B\oe hm \orcidB{}}
\email{celine.boehm@sydney.edu.au}
\affiliation{School of Physics, The University of Sydney and ARC Centre of Excellence for Dark Matter Particle Physics, NSW 2006, Australia}
\affiliation{The University of Edinburgh, School of Physics and Astronomy, EH9 3FD Edinburgh, UK}

\date{\today}


\begin{abstract}
Current ton-scale dark matter direct detection experiments have reached an important milestone with the detection of solar neutrinos. In this paper, we show that these data can be used to determine a critical parameter of the Standard Model in particle physics, across an energy regime that has never been probed before.  In particular, we show that the value of the weak mixing angle ($\theta_W$) which relates the mass of the $W$ and $Z$ bosons can be derived  from 1) the recent measurements of  coherent neutrino-nucleus scattering by PandaX-4T and XENONnT in the sub-GeV energy range -- a regime which is usually only probed by low energy neutrino experiments -- and from 2)  XENONnT electron recoil data through neutrino-electron scattering at energy scale $\simeq 0.1 ~ \rm{ MeV}$, corresponding to a momentum transfer region over an order of magnitude smaller than that explored by atomic parity violation experiments. Now that an indicative measurement of the weak mixing angle exists at these lowest energy frontier, the challenge for the next generation of such experiments will be to provide a more precise measurement in the keV-MeV energy range. 
\end{abstract}

\maketitle

\section{Introduction}
\label{sec:introduction}  
The proposal for the search of dark matter (DM) particles using direct detection (DD) experiments \cite{Goodman:1984dc} was initially inspired by the potential observation of MeV-range neutrinos through coherent neutrino-nucleus scattering (\cevns) \cite{Drukier:1984vhf}. Ironically, after decades of unfruitful searches for DM interactions with ordinary matter, current ton-scale DD experiments have started observing solar neutrinos through both \cevns \cite{PandaX:2024muv, XENON:2024ijk} and neutrino-electron scattering \cite{XENON:2022ltv,LZ:2023poo,PandaX:2024cic}, as anticipated in \cite{Drukier:1986tm, Monroe:2007xp, Strigari:2009bq, Gutlein:2010tq, Billard:2013qya, Gutlein:2014gma,Ruppin:2014bra, OHare:2016pjy, OHare:2021utq}. The observation of $^8$B solar neutrinos using \cevns has already  reached moderate statistical significance — $2.64\,\sigma$ and $2.73\,\sigma$ for PandaX-4T \cite{PandaX:2024muv} and XENONnT \cite{XENON:2024ijk}, respectively  and is likely to lead to a discovery in the future. The detection of neutrino-electron scattering has yet to achieve a similar statistical significance, even though $\mathcal{O}(10)$ solar neutrino events have been detected already \cite{XENON:2022ltv, LZ:2023poo, PandaX:2024cic}. However, the detection of these two types of scattering events by DM direct detection experiments represents a significant step-change for this technology, positioning it as a potential competitor to more conventional neutrino detectors. This paper explores whether these data can provide new insights into the Standard Model (SM) of Particle Physics. Interestingly, the answer turns out to be yes!

\begin{figure}
\centering
\includegraphics[width=\columnwidth]{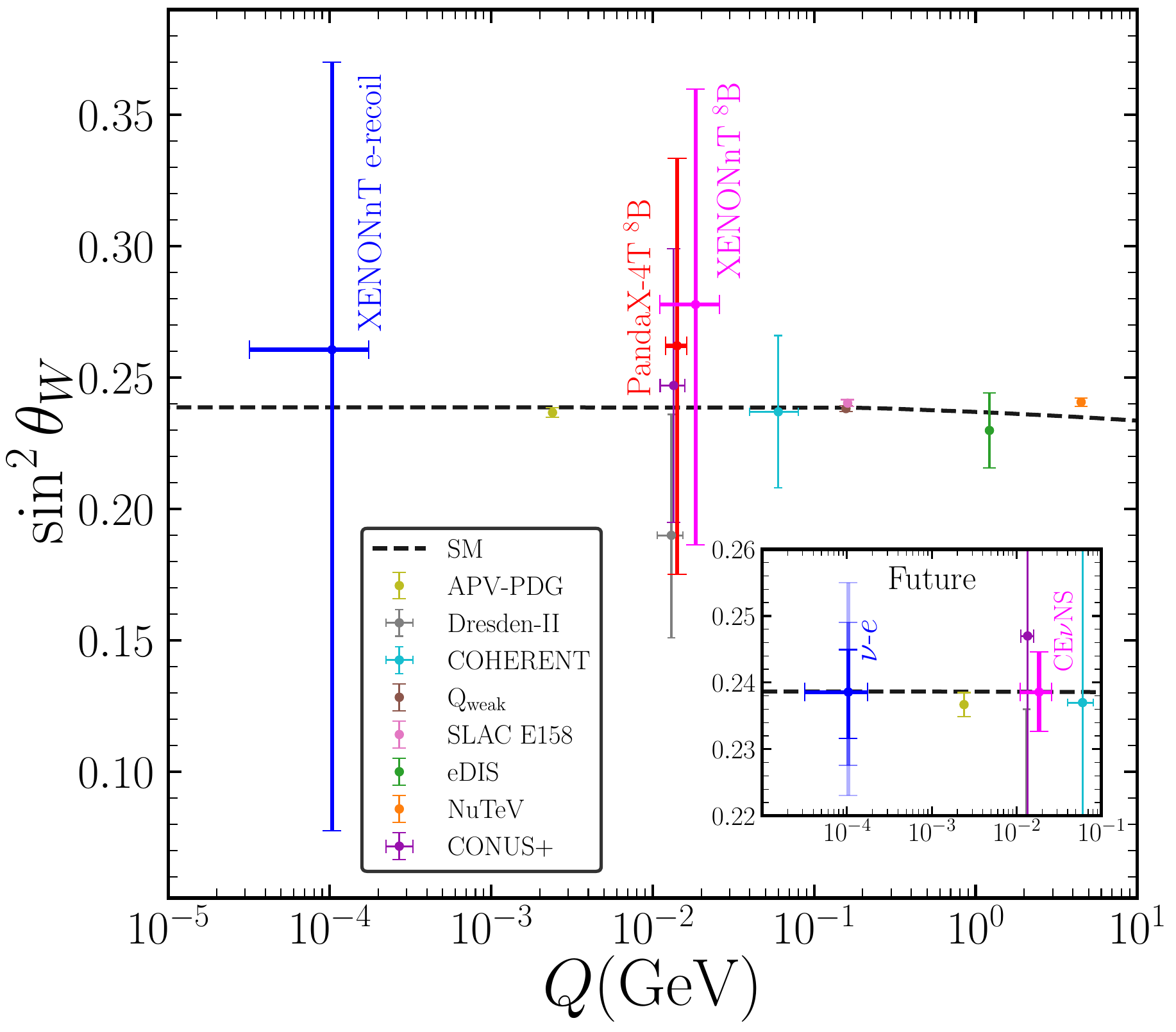}
\caption{Weak mixing angle as a function of the energy scale $Q$. Our $1\,\sigma$ measurements using the latest unpaired PandaX-4T $^8$B neutrino \cite{PandaX:2024muv}, XENONnT $^8$B neutrino \cite{XENON:2024ijk}, and XENONnT electron recoil \cite{XENON:2022ltv} data are shown in red, magenta, and blue, respectively. The SM prediction is represented by the dashed black line. Measurements from other considerations \cite{ParticleDataGroup:2022pth, Prescott:1979dh, SLACE158:2005uay, Qweak:2018tjf, COHERENT:2017ipa, COHERENT:2020iec, DeRomeri:2022twg, Colaresi:2022obx, AristizabalSierra:2022axl, Alpizar-Venegas:2025wor,DeRomeri:2025csu} are also shown by thin lines. In the inset, we show the future projection for a 200 ton-year exposure, with details provided in the Appendix.}
\label{fig:main} 
\end{figure}

In the SM, neutrinos interact through weak forces\,\cite{Weinberg:1967tq}. \cevns arises from the interactions of neutrinos with quarks via neutral current $Z$-mediated processes\,\cite{Freedman:1973yd}. The momentum transfer induced by  neutrinos in \cevns interactions is small enough that the corresponding de Broglie wavelength is larger than the typical nuclear radius. As a result, neutrinos perceive the nucleus as a whole, leading to a coherently enhanced cross section. The situation is different in the case of neutrino-electron scattering. First, in addition to the neutral current ($Z$-mediated) interactions, there is an additional contribution from the charged current ($W$-mediated) process\,\cite{Sarantakos:1982bp}. Second, since the electron is a point particle, there is no coherence effect in neutrino-electron scattering. Both of these processes depend on the weak mixing angle, $\theta_W$, a parameter that describes the mixing between the gauge boson of $U(1)_Y$ and the third component of the $SU(2)_L$ gauge boson. The $\theta_W$ parameter is related to the gauge couplings $g$ for $SU(2)_L$ and $g^\prime$ for $U(1)_Y$ through $\sin^2 \theta_W = {g^\prime}^2/(g^2 + {g^\prime}^2)$. The renormalization group equation \cite{Wilson:1971bg} indicates that the value of the gauge coupling depends on the energy scale, and so does the weak mixing angle. Over the decades, $\theta_W$ has been measured by many different experiments across various energy scales using different physical processes, such as atomic parity violation (APV) \cite{ParticleDataGroup:2022pth}, electron-deuteron deep inelastic scattering (eDIS) \cite{Prescott:1979dh, ParticleDataGroup:2022pth}, polarized Møller scattering (by SLAC E158) \cite{SLACE158:2005uay, ParticleDataGroup:2022pth}, elastic electron-proton scattering (by Q$_{\rm weak}$) \cite{Qweak:2018tjf, ParticleDataGroup:2022pth}, and neutrino scattering (by COHERENT\,\cite{COHERENT:2017ipa, COHERENT:2020iec, DeRomeri:2022twg}, Dresden-II \cite{Colaresi:2022obx, AristizabalSierra:2022axl}, CONUS+ \cite{Ackermann:2025obx, Alpizar-Venegas:2025wor,DeRomeri:2025csu}). The lowest energy probe among these measurements corresponds to the one obtained by the APV experiment, around 3 MeV. Here, we show that the detection of solar neutrinos by current DD experiments can be used to  measure $\sin^2 \theta_W$ in a momentum transfer regime which is an order of magnitude smaller than APV, thus making it the lowest energy probe achieved so far.

Current Xenon (Xe)-based DD experiments use a two-phase time projection chamber, consisting of liquid and gas phases, to detect potential DM events. An energy deposition in liquid Xe results in atomic motion which produces some unmeasurable heat, excitation, and ionization. Excitation leads to the emission of scintillation photons, observed as the S1 signal, while ionization leads to the S2 signal. Electron recoils are expected to produce more ionization than nuclear recoils. Therefore, in the $\sim$\keV scale recoil energy regime, these experiments can efficiently discriminate between nuclear and electron recoil events by comparing the S2/S1 ratio\,\cite{Aprile:2006kx}. This unique feature enables these experiments to search for new physics in both nuclear and electron recoil scenarios. However, below the $\sim$\keV scale recoil energy regime, the smallness of the S1 signal leads to a focus on S2-only analysis\,\cite{Essig:2012yx, XENON:2019gfn, PandaX:2022xqx}. An S2-only analysis loses the experimental capability to differentiate between nuclear and electron recoils due to the untraceable S2/S1 ratio\,\cite{Essig:2018tss, Wyenberg:2018eyv, Herrera:2023xun, Carew:2023qrj}.

The measurement of solar $^8$B neutrinos through \cevns by XENONnT was performed using a S1-S2 analysis (paired) \cite{XENON:2024ijk}, whereas PandaX-4T conducted the same measurement using both S1-S2 and S2-only analyses (unpaired)\,\cite{PandaX:2024muv}. In the PandaX-4T unpaired analysis, the contamination from neutrino-electron events is very small owing to its small cross section. We utilized PandaX-4T unpaired and XENONnT data to estimate $\sin^2 \theta_W$ using \cevns and the corresponding best fit values at $1\,\sigma$ are shown by the red and magenta data points in Fig.\,\ref{fig:main}. Clearly, current DD data not only provides  complementary results compared to  neutrino experiments but does so in a different momentum transfer regime. While the energy threshold of $^8$B solar neutrino search is low, the heavy Xe nuclear mass shifts the momentum transfer to the $\sim 10$\,MeV regime. This suggests that electrons  would be a better target to probe $\sin^2 \theta_W$ in the lowest  momentum transfer regime. This prompts us to use the latest XENONnT electron recoil results \cite{XENON:2022ltv} to find the best fit value for $\sin^2 \theta_W$. The corresponding result at $1\,\sigma$ is shown by the blue point in Fig.\,\ref{fig:main}. Remarkably, electron recoil events of XENONnT is probing $\sin^2 \theta_W$ at the lowest energy scale, an order of magnitude smaller than the APV measurement. Any DD experiment observing neutrino-electron scattering can achieve this, which implies that our work broadens the horizon of all DD experiments, enabling them to test the SM in an uncharted domain. We also show that future Xe-based DD experiments can measure $\sin^2 \theta_W$ with the precision of the percent level. With many planned DD experiments \cite{Akerib:2022ort}, a precise measurement of $\sin^2 \theta_W$ in these unexplored regimes may potentially indicate the presence of new physics. Such new physics arises in a broad class of theoretical models, primarily involving a new light mediator (e.g., see \cite{Davoudiasl:2014kua, Cadeddu:2021dqx, Davoudiasl:2023cnc}).

\section{Neutrino event rate}
\label{sec:nuRate}
In this section, we briefly discuss the neutrino-induced event rate following \cite{Billard:2013qya}. In our analysis, the source of the neutrinos is the Sun, as it produces neutrinos with the desired flux and energy. The neutrino-induced event rate is given by \cite{Billard:2013qya}
\begin{align}
\dv{R}{E_i} &= N_T \int_{E^{\text{min}}_{\nu,i}}\dv{\sigma}{E_{i}}\dv{\phi}{E_\nu}\mathrm{d}E_\nu \, ,
\label{eq:diffrate}
\end{align}
where $i \in {N,e}$ for nuclear and electron recoil respectively, $N_T$ is the number of target particles, $E_\nu$ refers to neutrino energy.  The solar neutrino fluxes (d$\phi/$d${E_\nu}$) and related uncertainties are adapted from \cite{OHare:2020lva}. The differential \cevns or $\nu$-$e$ cross section is represented by d$\sigma/$d$E_{i}$\footnote{We use tree level cross sections, see Refs.\,\cite{Marciano:1980pb, Bahcall:1995mm, Erler:2013xha, Tomalak:2020zfh} for the effect of radiative corrections.}. The minimum required neutrino energies $E^{\text{min}}_{\nu,i}$ for nuclear and electron recoil are \keep{:}
\begin{equation}
E^{\text{min}}_{\nu,N} = \sqrt{\frac{m_N E_N}{2}}; ~   E^{\text{min}}_{\nu,e} = \frac{E_e + \sqrt{E_e(E_e + 2m_e)}}{2}, 
\end{equation}
where $m_N$ and $m_e$ are the masses of the nucleus and the electron, respectively. The corresponding nuclear and electron recoil energies are $E_N$ and $E_e$, respectively. For the case of nuclear recoil the differential \cevns cross section is
\begin{align}
\dv{\sigma}{E_N} &= \frac{G_F^2}{4\pi}Q_W^2 m_N \left(1-\frac{m_N E_N}{2E_\nu^2}\right)F^2(E_N) \, ,
\label{eq:CEnuNS_xsection}
\end{align}
where the Fermi coupling constant $G_F = 1.166\times 10^{-5}$\,GeV$^{-2}$, and $F(E_{N})$ is the weak nuclear form factor, which generically depends on the nuclear recoil energy and the nuclear charge radius. However, in our analysis, we have assumed it to be the Helm form factor. In our recoil energy regime, the uncertainties related to the form factor are numerically insignificant. For a nucleus having $Z$ protons and $N$ neutrons, the weak nuclear hypercharge, $Q_W$, related to weak mixing angle through 
\begin{equation}
Q_W = N - Z(1-4\sin^2\theta_W)
\label{eq:nu_N_coupling}
\end{equation}
Unlike the tree level \cevns cross-section, the neutrino-electron scattering cross section is flavour dependent, and is given by
\begin{align}
\dv{\sigma_{\nu_l}}{E_e} = & Z_{\rm eff}^{\rm Xe}\left(E_e\right) \frac{G_F^2 m_e}{2\pi} \left[ \left(g_V^{\nu_l} + g_A^{\nu_l}\right)^2 +  \left(g_V^{\nu_l} - g_A^{\nu_l}\right)^2  \right. \nonumber\\
& \left.  \left(1- \frac{E_e}{E_\nu} \right)^2 - \left({g_V^{\nu_l}}^2 - {g_A^{\nu_l}}^2\right) \frac{m_e E_e}{E^2_{\nu}} \right]\, ,
\end{align}
where $Z_{\rm eff}^{\rm Xe}$ is the recoil energy dependent effective electron charge of Xe, adapted from \cite{Chen:2016eab, AtzoriCorona:2022jeb} and $m_e$ is the electron mass. The neutrino flavour specific vector (which depends on the weak mixing angle) and axial couplings to electrons are respectively
\begin{equation}
g_V^{\nu_l} =  2\sin^2\theta_W - \frac{1}{2} + \delta_{le}\,, \quad g_A^{\nu_l} = - \frac{1}{2} + \delta_{le}\, ,
\label{eq:nu_e_coupling}
\end{equation}
where the Kronecker delta function $\delta_{le}$ accounts for the effect of charged current interaction in  $\nu_e-e^-$ scattering. Finally, including the effect of neutrino oscillation, the total neutrino-electron cross section is
\begin{equation}
\dv{\sigma}{E_e} = P_{ee} \dv{\sigma_{\nu_e}}{E_e} + \sum_{l = \mu, \tau} P_{el} \dv{\sigma_{\nu_l}}{E_e}.
\label{eq:nu_e_xsection}
\end{equation}
The survival probability of $\nu_e$ is $P_{ee}$. The conversion probabilities of $\nu_e$ to $\nu_\mu$  and $\nu_\tau$ are denoted by $P_{e\mu}$ and $P_{e\tau}$. These probabilities depend on neutrino mixing angles \cite{Goswami:2004cn}, which are taken from Ref.\,\cite{ParticleDataGroup:2020ssz}, assuming normal ordering. Here we have assumed that $\sin^2\theta_W$ is independent of the transfer momentum, which is consistent with SM expectation \cite{Erler:2004in, Erler:2017knj}, for the range of interest.

\section{Analysis \& Results}
\label{sec:ana}
Building on the theoretical event rates discussed in the previous section, we now describe how we infer the value of the weak mixing angle using current DD results. While we primarily focus on Xe-based experiments, our analysis is generally applicable to most DD experiments. The analysis is divided into two parts: nuclear recoil and electron recoil. 

{\bf Nuclear recoil:} 
In this case, neutrinos coherently scatter off the nucleus of the target material. As mentioned earlier, DD experiments have already started observing these events at more than $2.5\,\sigma$. The \cevns is searched for in two ways: (i) using both S1 and S2 signals (paired) and (ii) using S2-only analysis (unpaired). The paired search is relatively clean but comes with a higher energy threshold. While XENONnT \cite{XENON:2024ijk} and PandaX-4T \cite{PandaX:2024muv} have observed $^8$B solar neutrinos using this method, there is no energy spectrum information available yet for PandaX-4T.  In contrast, the unpaired search has only been conducted by PandaX-4T. The unpaired signal is generated by ionized electrons accelerated through the electric field. Thus, even a small energy deposition can be amplified by the electric field to produce an observable signal. This results in a lower energy threshold compared to the paired search but at the cost of a larger background. Due to the lower threshold, the number of observed events is relatively high. For instance, in PandaX-4T, the number of best-fit $^8$B signal events obtained using the combined analysis for the paired and unpaired data samples are 3.5 and 75, respectively \cite{PandaX:2024muv}. We thus utilized the paired XENONnT and the unpaired PandaX-4T data sample to measure $\sin^2 \theta_W$. 

Given the \cevns differential event rate in Eq.\,\eqref{eq:diffrate} as a function of energy, we convert it into a differential event rate as a function of the number of electrons ($n_e$) for the unpaired data sample of PandaX-4T using
\begin{equation}
\dv{R}{n_e} = \mathcal{E} \times \dv{R}{E_N} \times\frac{1}{Q_y + E_N \dv{Q_y}{E_N}} \times {\rm efficiency}.
\label{eq:panda_rate}
\end{equation}
The PandaX-4T exposure ($\mathcal{E}$) is $1.04$ ton-year. For charge yield, $Q_y$,  we use the best-fit model of the same given in Fig.\,4 of Ref.\,\cite{PandaX:2024muv}. We have also used the selection efficiency from Fig.\,1 of Ref.\,\cite{PandaX:2024muv}, as the region of interest efficiency is already included in the charge yield. This approach reproduces the PandaX-4T $^8$B event rate appreciably, with a difference in the best-fit event rate of $\sim 10\%$.

For the XENONnT $^8$B data, we utilized the top panel of Fig.\,2 in Ref.\,\cite{XENON:2024ijk}. The event rate for each corrected S2 (cS2) bin is calculated using  
\begin{equation}
R = \mathcal{E}^{\prime} \int {\rm d(cS2)} \int {\rm d}{E_N} \, \dv{R}{E_N} \, \epsilon(E_N)  \, {\rm pdf}({\rm cS2}|E_N),
\label{eq:XenT_8B_rate}
\end{equation}
where the exposure, $\mathcal{E}^{\prime}$, is 3.51 ton-years. The energy-dependent acceptance, $\epsilon(E_N)$, is obtained from Fig.\,1 of \cite{XENON:2024ijk}. Following \cite{Szydagis:2021hfh}, we translate recoil energy to cS2. We have assumed a normalized Gaussian PDF to obtain the probability using the charge yield from \cite{XENON:2019izt} with the standard deviation derived from the error in electron gain ($g2$), quoted in Ref.\,\cite{XENON:2024ijk}. The cS2 binning for the SR0 and SR1 runs of the XENONnT $^8$B data are slightly different. We have used the average of these two binnings in our analysis. Using either the SR1 binning or the SR0 binning individually would change our best-fit value by $\sim$1\%.
In our numerical analysis, we employ the profile likelihood ratio test statistic \cite{Cowan:2010js, Baxter:2021pqo}
\begin{equation}
\label{eq:chisquare}
q_0 = -2\ln{\left[\frac{\mathcal{L}(\hat{\hat{\boldsymbol{\theta}}}|\mathcal{M}_{\nu + b})}{\mathcal{L}(\hat{\boldsymbol{\theta}}|\mathcal{M}_b)}\right]}, 
\end{equation}
where $\mathcal{M}_{b}$ represents the best fit model, and $\mathcal{M}_{\nu + b}$ represents the model combining both the signal (neutrinos, in our case) and the background. The nuisance parameter, $\boldsymbol{\theta}$, accounts for uncertainties in the relevant backgrounds for the background-only likelihood, $\mathcal{L}(\boldsymbol{\theta}|\mathcal{M}_b)$,  and both the uncertainties in the neutrino fluxes ($\phi^j$) and backgrounds for the combined one. The best fit and combined likelihood maximised at $\hat{\boldsymbol{\theta}}$ and $\hat{\hat{\boldsymbol{\theta}}}$, respectively. Note that $q_0$ follows a $\chi^2$ distribution. The combined likelihood is obtained using
\begin{equation}
\mathcal{L}(\boldsymbol{\theta}|\mathcal{M}_{\nu + b})= {\small \prod_{ i=1}^{ n}} \mathcal{P}(D^i {\Huge|}{\small \sum_{j=1}^{n_{\nu}}} N_{\nu}^i(\phi^j) + N_b^i){\small \prod_{\small k=1}^{ n_{\nu+b}}}\mathcal{G}(\theta^k).
\label{eq:llhood}
\end{equation}
Here $\mathcal{P}$ denotes the Poisson probability. The Gaussian distributions, $\mathcal{G}$, account for uncertainties in the neutrino fluxes and backgrounds. The background rate and the data in the $i^{\text{th}}$ bin are represented by $N_b^i$ and $D^i$, respectively. The quantity $N^i_{\nu}(\phi^j)$ represents events generated by $j^{\text{th}}$ type solar neutrinos in the $i^{\text{th}}$ bin. The total number of solar neutrinos and background contributions is $n_{\nu + b}$, while for neutrinos alone, it is $n_{\nu}$. The maximum number of bins included in the analysis is $n$. The background only likelihood can be obtained excluding the contribution of neutrinos from Eq.\,\eqref{eq:llhood}.

%
%
\begin{figure}
\centering
\includegraphics[width=\columnwidth]{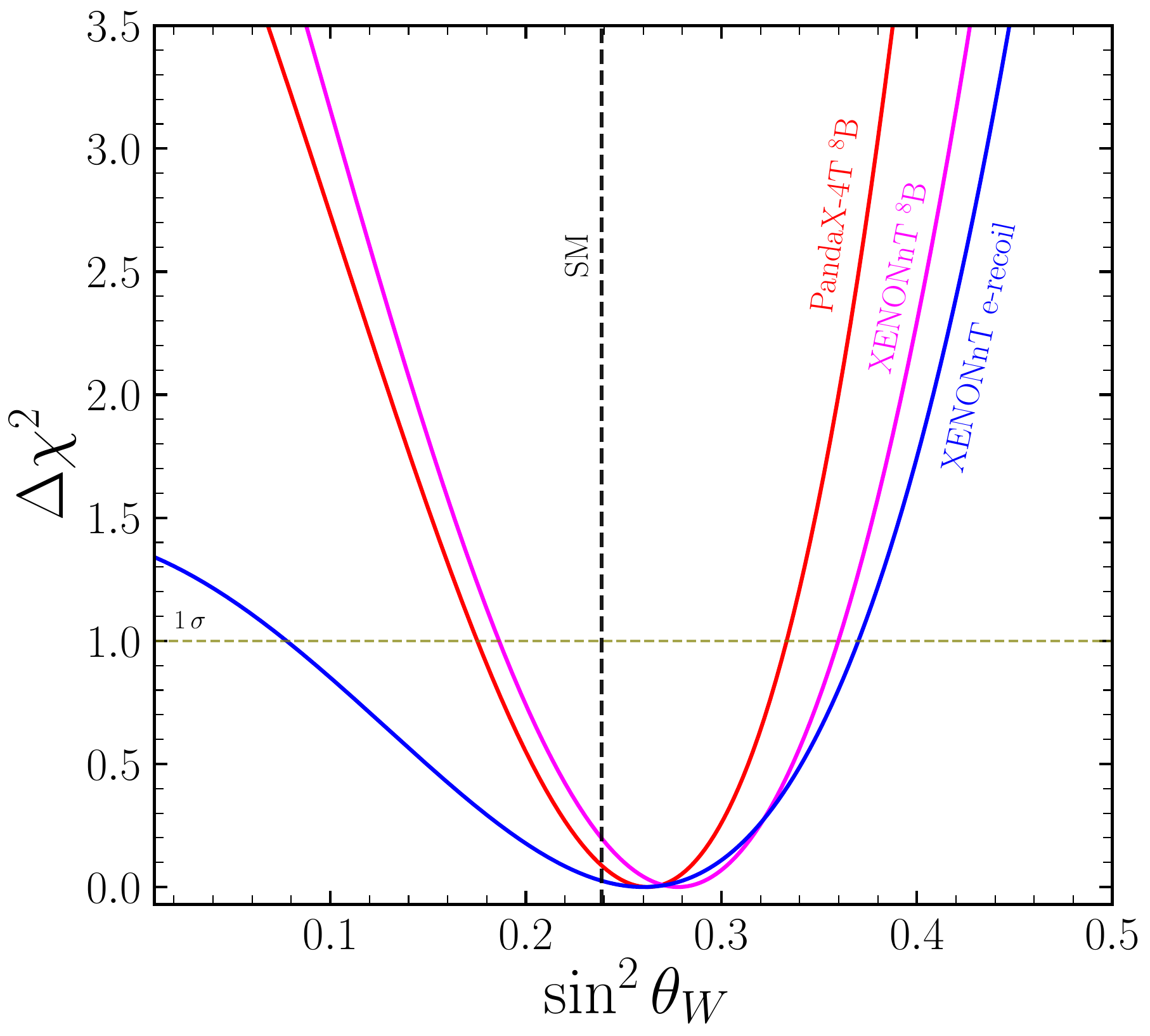}
\caption{The variation of $\Delta \chi^2$ with $\sin^2\theta_W$. The red, magenta and blue solid lines correspond to the latest unpaired PandaX-4T $^8$B solar neutrino, XENONnT $^8$B solar neutrino and XENONnT electron recoil data samples, respectively. The SM prediction for very low momentum transfer is shown by the dashed black line. Our $1\,\sigma$ limits on $\sin^2\theta_W$ can be followed from the dashed olive line.}
\label{fig:del_chisquare} 
\end{figure}

For PandaX-4T, the $^8$B neutrino-induced rate $N^i_{\nu}$ can be evaluated from Eq.\,\eqref{eq:panda_rate}. The uncertainties in the background rates are obtained from Table III of  \cite{PandaX:2024muv}. Like PandaX-4T \cite{PandaX:2024muv}, we have included only the first $8$ bins (i.e., $n_e = 4$ to $n_e = 8$) in our analysis. The corresponding $\Delta \chi^2$ against $\sin^2\theta_W$ is displayed by the red solid line in Fig.\,\ref{fig:del_chisquare}, labelled as PandaX-4T $^8$B. For the XENONnT $^8$B data, neutrino generated events are calculated using Eq.\,\eqref{eq:XenT_8B_rate} and the background uncertainties are adopted from Table I of Ref.\,\cite{XENON:2024ijk}. The associated $\Delta \chi^2$ is displayed by the magenta line in Fig.\,\ref{fig:del_chisquare}.  Remarkably, in both analyses, the best-fit value is close to the SM prediction, indicated by the dashed black line in Fig.\,\ref{fig:del_chisquare}. The best-fit values of $\sin^2\theta_W$ at $1\,\sigma$ for PandaX-4T $^8$B unpaired data and XENONnT $^8$B data  are depicted by the red and magenta points respectively in Fig.\,\ref{fig:main}.  The SM prediction against $Q$ is shown by the dashed black line in Fig.\,\ref{fig:main}. Further, we have displayed results from various other experiments including the results using dedicated neutrino experiments, such as COHERENT \cite{DeRomeri:2022twg} and DRESDEN-II \cite{AristizabalSierra:2022axl}, which lie in a similar momentum transfer regime\footnote{Please see Refs.\,\cite{Reines:1976pv, Boehm:2004uq, Khan:2016uon, Canas:2016vxp, Khan:2017oxw, Khan:2017djo, Canas:2018rng, Borexino:2019mhy, deGouvea:2019wav,   Miranda:2020tif, Cadeddu:2021ijh, COHERENT:2021xmm, Majumdar:2022nby, AtzoriCorona:2023ktl, AtzoriCorona:2024vhj, Chen:2024tqh, DeRomeri:2024iaw} for other similar searches.}.  Our results probe $\sin^2\theta_W$ in a different momentum transfer regime.

We stress that while numerous studies explore the prospect of probing beyond the SM physics using \cevns at future and current DD \cite{Harnik:2012ni, Cerdeno:2016sfi, Bertuzzo:2017tuf, Dutta:2017nht, Gonzalez-Garcia:2018dep,  Boehm:2018sux, Huang:2018nxj, Link:2019pbm, Boehm:2020ltd, Amaral:2020tga, AristizabalSierra:2020edu, Khan:2020vaf, Karmakar:2020rbi, Seto:2020udg, Khan:2020csx, Majumdar:2021vdw, Li:2022jfl, Khan:2022bel, A:2022acy, Amaral:2023tbs, Giunti:2023yha, Demirci:2023tui,  DeRomeri:2024dbv, AristizabalSierra:2024nwf, Herrera:2024ysj}, to the best of our knowledge, this is the first study to probe a SM parameter using current DD data. We provide the numerical value of $\sin^2\theta_W$ for PandaX-4T analysis below
\begin{equation}
\sin^2\theta_W = 0.26_{-0.09}^{+0.07}\left(1\,\sigma\right)_{-0.16}^{+0.11}\left(90\%\,{\rm CL}\right). 
\end{equation}
The quoted values are for the momentum transfer range $[0.012-0.016]$\,\GeV, which is determined by the recoil energy regime of PandaX-4T's unpaired $^8$B data sample. Our estimate of \( \sin^2 \theta_W \), obtained from the PandaX-4T data, lies above the fiducial SM value (similar to Ref.\,\cite{DeRomeri:2024iaw}) and is in agreement with the trend indicated by their results\,\cite{PandaX:2024muv}.

For XENONnT $^8$B  dataset, $\sin^2\theta_W$ is
\begin{equation}
\sin^2\theta_W = 0.28_{-0.09}^{+0.08}\left(1\,\sigma\right)_{-0.16}^{+0.13}\left(90\%\,{\rm CL}\right). 
\end{equation}
The values mentioned above are for the momentum transfer range $[0.011-0.026]$\,\GeV. Our XENONnT best-fit value lies above the SM expectation because we could only use the top panel of Fig.\,2 from the XENONnT paper~\cite{XENON:2024ijk} in our analysis. In that plot, the data in the first energy bin lies above the SM prediction. Furthermore, the XENONnT collaboration employed a boosted decision tree (BDT) score to distinguish $^8\mathrm{B}$ \cevns events from background, which we did not attempt, as replicating their methodology with the limited available information is highly challenging.

Although the thresholds of the aforesaid analysis are low, the heavy Xe nucleus drives the momentum transfer to the $\sim 10$\,\MeV range. This implies that an electron recoil search would be an ideal setup to probe $\sin^2\theta_W$ at the lowest energy scale. We now turn to this discussion.

{\bf Electron recoil:} As mentioned earlier, Xe-based experiments can efficiently discriminate between nuclear and electron recoil by comparing the ratio of S2/S1 in the $\gtrsim$\,\keV recoil energy range. Thus a search for $\nu$-$e$ scattering using electron recoil data enables these experiments to measure $\sin^2\theta_W$\footnote{This has also been realised in Refs.\,\cite{Cerdeno:2016sfi, DARWIN:2020bnc, Aalbers:2022dzr}, however we used current data.}. We utilized the latest XENONnT electron recoil data sample in our analysis \footnote{We have not used the LZ \cite{LZ:2023poo} and PandaX-4T \cite{PandaX:2024cic} electron recoil data due to their lower sensitivity.}. The neutrino-induced electron recoil events are evaluated using Eq.\,\eqref{eq:diffrate} with the cross section given in Eq.\,\eqref{eq:nu_e_xsection}. The differential event rate with respect to the reconstructed energy ($E_e^{\rm res}$) is given by
\begin{equation}
\dv{R}{E_e^{\rm res}} = \int \dv{R}{E_e} \, \epsilon \left(E_e^{\rm res}\right) \, G(E_e^{\rm res},E_e,\sigma)\, {\rm d}{E_e},
\label{eq:nu_rate_XenT_erecoil}
\end{equation}
where $\epsilon \left(E_e^{\rm res}\right)$ is the total efficiency given in Fig.\,1 of \cite{XENON:2022ltv}. The event rate is smeared with a normalised Gaussian function, $G$, having energy resolution $\sigma$, stated in Ref.\,\cite{XENON:2020iwh}. In our statistical analysis we have again used Eq.\,\eqref{eq:chisquare} with $N^i_\nu$ obtained from Eq.\,\eqref{eq:nu_rate_XenT_erecoil}. The data $D_i$ is extracted from Ref.\,\cite{XENON:2022ltv}. The post-fit background rate provided for 1-30\,keV recoil in Ref.\,\cite{XENON:2022ltv} includes the SM $\nu$-$e$ rate. Since our analysis focuses on searching for $\nu$-$e$ scattering in the same data, our background model excludes this rate (and associated uncertainty), assuming the experiment used the expected low energy  SM value for $\sin^2\theta_W=0.2386$ \cite{Erler:2017knj}, to avoid double counting. We have also excluded first bin from data analysis as the efficiency falls below $10\%$ at energies $\lesssim 1\,$keV$_{ee}$, hence $n = 29$ in Eq.\,\eqref{eq:llhood}. 

The corresponding $\Delta \chi^2$ is depicted by the solid blue line in Fig.\,\ref{fig:del_chisquare}. The best fit value of $\sin^2\theta_W$  at $1\,\sigma$ is shown in Fig.\,\ref{fig:main} by the blue data point. Expectedly the error bar is rather large as the experiment itself has not observed $\nu$-$e$ scattering events with desirable significance.  We note that above $\sim 1.16\,\sigma$, we could only get an upper limit in the value of  $\sin^2\theta_W$. We now   present the numerical value of $\sin^2\theta_W$ from the electron recoil analysis.
\begin{equation}
\sin^2\theta_W = 0.26_{-0.18}^{+0.11}\left(1\,\sigma\right)^{+0.17}\left(90\%\,{\rm CL}\right). 
\end{equation}
The reported values correspond to a momentum transfer range of $[3.20 \times 10^{-5}-1.75 \times 10^{-4}]$\,\GeV. While the recoil energy regime of this analysis is similar to the nuclear one, the significant mass ratio between the Xe nucleus and the electron allows us to probe $\sin^2\theta_W$ in a momentum transfer region that has not been explored by any other experiments before. The closest comparison is with the APV result, which is in a momentum transfer regime more than an order of magnitude higher. Therefore,  even obtaining an upper limits using current data is a remarkable achievement for DD\footnote{We note that data of experiments like Borexino \cite{Borexino:2017rsf}, SNO$+$ \cite{SNO:2024vjl} could fill the gap between our XENONnT and APV results in Fig.\,\ref{fig:main}.}. Furthermore, these experiments are expected to improve their understanding of the electron recoil background in the near future. Notably, within two years, the XENONnT data sample \cite{XENON:2022ltv} has reduced background events by almost a factor of $5$ compared to the XENON1T electron recoil excess data sample \cite{XENON:2020iwh}. We have demonstrated the potential for future improvement using a 200 ton-year exposure for both \cevns and $\nu-e$ scattering in the inset of Fig.\,\ref{fig:main}, with relevant details provided in Appendix. As shown in the figure, future experiments can measure $\sin^2 \theta_W$ with approximately 2\% accuracy. This shows potential discovery of solar neutrino interactions would definitely reduce the error bars associated with our results \cite{PANDA-X:2024dlo, DARWIN:2020bnc, XLZD:2024nsu} and may indicate the possible presence of new physics. It is therefore worth making a dedicated effort to study the weak mixing angle in DM DD experiments, especially if one uses the electron recoil channel.
\section{Conclusions}
\label{sec:conclusions}

In this paper, we demonstrate that current DD data can be used to measure the weak mixing angle. We show that the latest $^8$B solar neutrino measurements from PandaX-4T and XENONnT can probe $\sin^2\theta_W$ in a region complementary to the dedicated neutrino experiments. Furthermore, we emphasize that electron recoil measurements can help to explore $\sin^2\theta_W$ in a completely new energy scale through neutrino-electron scattering. The current XENONnT electron recoil data already probe $\sin^2\theta_W$ in a momentum transfer region that is an order of magnitude smaller than that of the APV result. Our results agree with SM expectation; however, it is too early to draw conclusions about the possible presence of new physics given our error bars. We also estimate the potential improvement that these experiments could achieve with a $200$ ton-year exposure, which is feasible for experiments like PandaX-xT \cite{PANDA-X:2024dlo} and XLZD \cite{XLZD:2024nsu}. Our findings indicate that these experiments would significantly improve the precision of the measurement.

 While we have focused specifically on Xe-based experiments, our exploration is generically   applicable to all DD experiments. In the context of currently running experiments, DarkSide \cite{DarkSide-20k:2017zyg}  would be able to study $\sin^2\theta_W$ in a different region once it starts observing a significant number of neutrino events. The ability of such experiments  to discriminate between nuclear and electron recoil using pulse shape analysis would be particularly useful for investigating $\sin^2\theta_W$ in a previously unexplored region, similar to our result using the XENONnT electron recoil search. Proposed low-threshold DD experiments like Oscura \cite{Oscura:2022vmi} would  also be valuable for this purpose. In the future, if these low-threshold DD experiments can differentiate between electron and nuclear recoil and begin observing neutrino events then they may be able to probe the weak mixing angle in the lowest possible momentum transfer region due to their extremely low threshold. In summary, our work opens pathway to probe a SM parameter in a previously unexplored domain using DD experiments, thus offering a potential opportunity to discover new physics.

\emph{Acknowledgments --} We thank Qing Lin for useful correspondence regarding PandaX-4T $^8$B solar neutrino result. We also thank Theresa Fruth, Ranjan Laha and Ciaran O'hare for discussions. TNM thanks Debajit Bose for his help with the plot. The work of TNM is supported by the Australian Research Council through the ARC Centre of Excellence for Dark Matter Particle Physics.

\section{Appendix}
\label{sec:app}
In this Appendix, we discuss the possibility of measuring the weak mixing angle in future xenon-target experiments. For both electron and nuclear recoil channels, we assume a $200$ ton-year exposure. For the electron recoil channel via $\nu-e$ scattering, we consider the dominant electron recoil backgrounds from \cite{DARWIN:2020bnc,Newstead:2018muu}, namely, $^{136}\text{Xe}$, $^{124}\text{Xe}$, and Kr backgrounds with three choices background uncertainties $1\%$ \cite{Newstead:2018muu,deGouvea:2021ymm}, $3\%$, $6\%$. We incorporate the efficiency from XENONnT \cite{XENON:2024ijk} and restrict the analysis to the recoil energy range of $1-30$ keV. Along with the aforementioned backgrounds, we include the expected solar neutrino signal in the mock data, assuming SM expectation for $\sin^2 \theta_W$. We then apply the profile likelihood method, as described in the main text, to measure $\sin^2 \theta_W$. The corresponding results are shown in the inset of Fig.\,\ref{fig:main} by the blue solid lines  representing the three choices of background uncertainties mentioned above. Note that the measurement uncertainty increases with increasing background uncertainty, as indicated by the decreasing opacity of the blue lines. As expected, with a $200$ ton-year exposure, the error bar on the current measurement significantly decreases. For the nuclear recoil channel, we restrict the analysis to the recoil energy range of $0.5-3$ keV to evade uncertainties from the nuclear form factor. Additionally, we do not include any background contributions, as this search is expected to be less affected by backgrounds. The result using a $200$ ton-year exposure is shown in the inset of Fig.\,\ref{fig:main} by the magenta solid line, where we also observe a substantial improvement compared to current results.

\bibliographystyle{JHEP}
\bibliography{ref.bib}

\end{document}